\documentstyle[12pt,aaspp4,psfig,flushrt,tighten]{article}

\lefthead{Schneider, Guetta, Ferrara}
\righthead{High Energy Neutrinos from the First Stars}

\def\be{\begin{equation}}
\def\ee{\end{equation}}

\def\msun{{M_\odot}}

\def\etal{{\it et al.~}}

\def\gsim{\lower.5ex\hbox{\gtsima}}
\def\lsim{\lower.5ex\hbox{\ltsima}}
\def\gtsima{$\; \buildrel > \over \sim \;$}
\def\ltsima{$\; \buildrel < \over \sim \;$}
\def\prosima{$\; \buildrel \propto \over \sim \;$}
\def\gsim{\lower.5ex\hbox{\gtsima}}
\def\lsim{\lower.5ex\hbox{\ltsima}}
\def\simgt{\lower.5ex\hbox{\gtsima}}
\def\simlt{\lower.5ex\hbox{\ltsima}}
\def\simpr{\lower.5ex\hbox{\prosima}}

\def\ie{{\frenchspacing\it i.e. }}
\def\eg{{\frenchspacing\it e.g. }}

\def\SNgg{SN$_{\gamma\gamma}$}
\def\fgg{f_{\gamma\gamma}}

\def\beq#1{\begin{equation}\label{#1}}
\def\eeq{\end{equation}}
\def\beqa#1{\begin{eqnarray}\label{#1}}
\def\eeqa{\end{eqnarray}}

\def\H2p{H$_2^+$ }

\def\mH2p{H_2^+}

\input psfig
\begin{document}

\title{Gamma Ray Bursts from the First Stars: Neutrino Signals}

\author{Raffaella Schneider, Dafne Guetta, Andrea Ferrara}

\affil{\footnotesize Osservatorio Astrofisico di Arcetri, Largo Enrico Fermi 5, 
	  50125 Firenze, Italy}

\begin{abstract}

If the first (PopIII) stars were very massive, their final fate is to 
collapse into very massive black holes. 
Once a proto-black hole has formed into the stellar core, accretion 
continues through a disk. It is widely accepted, 
although not confirmed, that magnetic fields drive an energetic 
jet which produces a burst of TeV neutrinos by photon-meson interaction,
and eventually breaks out of the stellar envelope appearing as a
Gamma Ray Burst (GRB). Based on recent numerical 
simulations and neutrino emission models, we predict the expected neutrino 
diffuse flux from these PopIII GRBs and compare it with the 
capabilities of present and planned detectors as AMANDA and IceCube.
If beamed into 1\% of the sky, we find that the rate of PopIII GRBs is 
$\le 4 \times 10^6$~yr$^{-1}$.
High energy neutrinos from PopIII GRBs could dominate 
the overall flux in two energy bands [$10^4 - 10^5$] GeV and 
[$10^5 - 10^6$] GeV of neutrino telescopes. 
The enhanced sensitivities of forthcoming 
detectors in the high-energy band (AMANDA-II, IceCube) will provide a fundamental insight on the characteristic 
explosion energies of PopIII GRBs and will constitute a unique probe of the 
the Initial Mass Function (IMF)  of the first stars and of the redshift $z_f$ marking the 
metallicity-driven transition from a top-heavy to a normal IMF.
The current upper limit set by AMANDA-B10 implies that such transition must have occurred
not later than $z_f =9.2$ for the most plausible jet energies. 
Based on such results, we 
speculate that PopIII GRBs, if not chocked, could be associated
with a new class of events detected by BeppoSax, the Fast X-ray Transient (FXTs), 
which are bright X-ray sources, with peak energies
in the $2-10$~keV band and durations between $10-200$~s.

\end{abstract}


\section{Introduction}

One of the most challenging problems in modern cosmology is the 
understanding of the first episodes of star formation in the Universe.
A growing body of theoretical work has been devoted in the
past few years to this subject (Rees 1976; Rees \&
Ostriker 1977; Silk 1977, 1983; Haiman, Thoul \& Loeb 1996; Uehara
\etal 1996). These investigations are aimed at the 
identification of the characteristic mass scale of the first
stars (usually referred to as PopIII stars), how this relates to the
physical conditions of the gas in the first collapsed objects and how
it affects subsequent structure formation in the Universe.

Numerical simulations (Abel \etal 1998; Nakamura \& Umemura 1999;
Bromm, Coppi \& Larson 1999, 2001; Abel, Bryan \& Norman 2000; 
Bromm \etal 2001; Ripamonti \etal 2001) based on hierarchical 
scenarios of structure formation and/or detailed stellar collapse models
 have shown that the typical mass scale for the first collapsed
clumps of primordial gas is $\approx 10^{3} \msun$, which corresponds to the 
Jeans mass set by molecular hydrogen cooling. 

The ultimate nature of the stars that form out of these clumps critically
depends on the physical conditions of the gas as the 
evolution pushes density to higher values. 
Preliminary studies (Omukai \& Nishi 1998; Nakamura \& Umemura 1999; Bromm \etal 2001;  
Ripamonti \etal 2001) show that as long as the metallicity is
below some critical value (typically $Z_{\rm cr} \sim 10^{-4} Z_{\odot}$) 
the first clumps have little tendency to further fragment, as 
also expected from a number of physical arguments (see \eg Schneider \etal 2001).
 
Tentative evidences for an early top-heavy initial mass function (IMF) are provided by a number of
observations (see Hernandez \& Ferrara 2000 and references therein).
For instance, a comparison between the observed 
number of metal poor stars with the one predicted by cosmological models,
implies that the stellar characteristic mass sharply increases with 
redshift. Furthermore, the
intracluster medium (ICM) metal abundances measured from {\it Chandra} 
and {\it XMM} spectral data are
higher than expected from the enrichment by standard IMF supernova (SN) yields
in cluster galaxy members, indicative 
of a top-heavy early IMF.  Finally, the observed
abundance anomalies (\eg oxygen) in the ICM can be 
explained by an early generation of PopIII SNe (Loewenstein
2001). 

These issues, highly suggestive of a top-heavy early star
formation, have recently motivated a series of numerical investigations
of the nucleosynthesis and final fate of metal-free massive stars
(Heger \& Woosley 2001; Fryer, Woosley \& Heger 2001). Stars with masses
in the range $140 \msun \leq M \leq 260 \msun$ undergo electron-positron
pair instability  (\SNgg) and end up in a giant, nuclear-powered explosion, leaving
no remnant and enriching the ambient medium with their nucleosynthetic products.
The kinetic energy released during the thermonuclear explosions 
powered by pair instability are $\approx 10^2$ times larger than those of ordinary Type II
SNe. This might cause the interaction with the circumstellar medium to be
as strong as predicted for hypernovae (Woosley \& Weaver 1982).
These explosions do not lead to the ejection of strongly
relativistic matter and therefore cannot power a Gamma-Ray Burst
(GRB, Fryer, Woosley \& Heger 2001).
 
However, if stars have masses $M>260 \msun$, photodisintegration 
instability is encountered before explosive nuclear burning can reverse
the implosion and the stars collapse to Very Massive Black Holes (VMBHs),
swallowing virtually all previously produced heavy elements. 
These stars are likely to be rapidly
rotating and the estimated angular momentum is sufficient to delay
black hole formation (Fryer, Woosley \& Heger 2001). Once a proto-black hole
has formed into the core, accretion continues through a disk at a rate
which can be as large as $1-10\, \msun \, {\rm s}^{-1}$.
It is widely accepted, although not confirmed, that 
magnetic fields might drive an energetic jet which can produce
a strong GRB through the interaction with surrounding gas.

In this scenario, the energetic jets generated by GRBs engines produce,
by photon-meson interaction, a burst of TeV neutrinos while propagating
in the stellar envelope (M{\'e}sz{\'a}ros \& Waxman 2001). It is widely
assumed that if the progenitors of GRBs are massive, collapsing stars
(the so called ``collapsar'', Woosley 1993, Paczy\'nski 1998) the shocks
producing the $\gamma$-rays occur after the relativistic jet has emerged 
from the stellar envelope. For a significant fraction of collapsars, 
the jet may be unable to punch through the stellar envelope (MacFadyen, Woosley \& Heger 2001).
However, the TeV neutrino signal from such ``chocked" jets should be similar to
that from jets which do break through the stellar envelope, leading to observable
GRBs. Detecting the neutrino flux from individual collapsars may provide a direct
evidence of such $\gamma$-ray-dark collapses (M{\'e}sz{\'a}ros \& Waxman 2001).

In this paper, we compute the diffuse flux of high energy neutrinos emitted
by PopIII GRBs at high redshifts ($5.4 < z < 30$). Hereafter, by PopIII GRBs
we indicate PopIII stars with masses $>260 \msun$ which collapse to a VMBHs
after a transient phase of mass accretion. 

Following Schneider \etal 2001, we assume that a fraction $(1-\fgg)$ of
the first stars collapse to VMBHs powering a relativistic jet that generates a flux of 
high energy neutrinos. Using the neutrino emission model proposed by 
M{\'e}sz{\'a}ros \& Waxman (2001) and integrating over the source rate throughout the universe,
we compute the diffuse neutrino flux from PopIII GRBs. 

The sensitivities of present (AMANDA\footnote{\tt
http://amanda.berkeley.edu/amanda/amanda.html}, Andr{\'e}s \etal 2001) and
forthcoming neutrino telescopes 
(such as AMANDA-II, Barwick 2001, IceCube\footnote{\tt http://www.ssec.wisc.edu/a3ri/icecube/}, Spiering 2001 
and other km-scale detectors, Halzen 2001)
enable to derive important constraints on the first episodes of star formation,
such as the relative number of \SNgg \, and VMBHs (\ie PopIII GRBs), metal enrichment
and the nature of dark matter.

The paper is organized as follows: in Section 2 we describe the proposed 
scenario for the formation of PopIII stars and we compute the expected rate of GRBs. 
In Section 3 we determine the energy spectrum of TeV neutrinos emitted in individual
PopIII GRBs. In Section 4 we compute the expected diffuse flux from PopIII GRBs.
In Section 5 we explore the parameter space making detailed predictions for
the detectability with present and forthcoming neutrino telescopes. In Section 6 we
illustrate the implications of the proposed scenario for PopIII GRBs.
Finally, in Section 7 we summarize our main results.

\section{Population III stars as GRB progenitors}

In this Section we discuss the formation and evolution of PopIII stars. 
We work within the paradigm of hierarchical cold dark matter (CDM) models
for structure formation, wherein dark matter halos collapse and the
baryons in them condense, cool and eventually form stars.
We adopt a cluster-normalized $\Lambda$CDM cosmological model with the following parameters: 
$\Omega_M=0.3$, $\Omega_{\Lambda}=0.7$, $h=0.65$ and $\Omega_B h^2=0.019$. 

If a fraction of the first stars are progenitors
of VMBHs and can power a GRBs, we expect a large number of TeV neutrinos
emission episodes to have occurred throughout the Universe. 
In order to estimate the diffuse flux contributed by this large ensemble
of uncorrelated neutrino bursts, we need to compute the expected number of sources
as a function of redshift. Following Schneider \etal 2001, to each halo of total mass $M$, 
we associate a baryonic mass given by $(\Omega_B/\Omega_M) M$. Here we consider only halos within which
gas can cool and form stars. In agreement with numerical simulations (Bromm \etal 2001), 
we assume that only a fraction $f_{\star} = 1/2$ of the baryonic gas 
turn into stars, the rest remaining in diffuse
form. The relative fraction of \SNgg \, and VMBH progenitors are
parametrized as follows:
\begin{eqnarray}
\label{eq:IMF}
M_{\gamma\gamma} & = & f_{\gamma\gamma} \frac{M}{2}\,(\frac{\Omega_B}{\Omega_M}) = f_{\gamma\gamma} M_*,\\ \nonumber
M_{\bullet}& = & (1-f_{\gamma\gamma}) \frac{M}{2}\,(\frac{\Omega_B}{\Omega_M}) = (1-f_{\gamma\gamma}) M_*,
\end{eqnarray}
where $M_{\gamma\gamma}$ ($M_{\bullet}$) is the total object mass which ends
up in \SNgg \,  (VMBHs) and $M_*$ is the mass processed into stars. Thus,
only a fraction $(1-\fgg)$ of the formed stars might generate a
relativistic jet during the VMBH collapse, leading to a TeV neutrino burst.
The remaining fraction $\fgg$ contribute to the metal enrichment of the 
intergalactic medium (IGM) and is responsible for the transition to a standard
fragmentation mode, hence IMF, which occurs when the average metallicity is
above the critical value $<Z_{\rm cr}> \, \approx 10^{-4} Z_{\odot}$ at 
redshift $z_f$.
Using the \SNgg \, metal yields given by Heger \& Woosley (2001) (see also Umeda \& Nomoto 2002) 
and assuming that metals pollute uniformly the IGM, Schneider \etal (2001) 
have inferred the transition redshift, $5.4 \le z_f \le 11$, from a top-heavy to a standard
IMF as a function of $\fgg$. These parameters directly affect the number 
of PopIII GRBs (or VMBHs) predicted by the model, as they define the duration and efficiency
of massive PopIII star formation epoch. Here we are interested to redshifts $z \gsim z_f$  
within a range $5.4 \le z \le 30$. 
If $\fgg \approx 1$, \ie all PopIII stars are \SNgg, metal enrichment is very efficient and transition
to a normal IMF occurs at $z_f \approx 11$. Conversely, as the relative fraction of \SNgg \, to VMBHs
decreases, massive PopIII star formation can proceed down to a minimum redshift $z_f = 5.4$ which 
corresponds to $\fgg^{\rm min} = 8 \times 10^{-3}$. At this epoch, the amount of VMBHs formed can 
account for the entire baryonic dark matter content of galaxy halos (see Schneider \etal 2001). 

As our reference model PopIII GRB progenitor, we consider a $300 \msun$ star. 
This choice is
motivated by the availability of detailed and complete evolutionary models of zero-metallicity 
stars of this mass (Fryer, Woosley \& Heger 2001). These simulations include
angular momentum transport through all burning stages until complete collapse to VMBHs
occurs. 

We compute the star formation rate of PopIII stars as,
\be
\Psi(z) = \frac{d}{dt} \int_{M_{\rm min}(z)}dM n(M,z)M_*
\ee
where $n(M,z)$ is the comoving number density of halos per unit mass predicted by the Press-Schechter
formalism, $M_*$ is given by eq.~\ref{eq:IMF} and the integration is performed from $M_{\rm min}(z)$
which is the minimum mass that can cool within a Hubble time at the specified formation redshift $z$,
\ie ${\rm t_{cool}}(M,z)\lsim {\rm t_{H}}$ (see Ciardi \etal 2000). Therefore, we assume that all gas
in halos with $M>M_{\rm min}(z)$ is able to form stars with an efficiency $f_\star$. 
 
The rate of PopIII GRBs produced by $300 \msun$ progenitor stars is
\be
\label{eq:rate}
R(z) = \frac{(1-\fgg)}{300 \msun}
\int_{z_f} dz \frac{\Psi(z)}{(1+z)} \frac{dV}{dz},
\ee   
where $dV$ is the comoving volume element,
\be
\label{vol}
dV = 4 \pi \left(\frac{c}{H_0}\right) \, r(z)^2 \, 
{\cal F}(\Omega_M,\Omega_{\Lambda},z) \, dz,
\ee
$H_0 = 100~h {\rm km}{\rm s}^{-1}{\rm Mpc}^{-1}$ is the Hubble constant, $r(z)$ is the comoving distance and 
\be
{\cal F}(\Omega_M,\Omega_{\Lambda}, z) \equiv 
\frac{1}{\sqrt{(1+z)^2\,(1+\Omega_M\,z)-z\,(2+z)\,\Omega_
{\Lambda}}}.
\ee
The results are shown in Fig.~1 (solid lines). The
maximum and minimum rates correspond to the extremes values for $\fgg$ (or equivalently, $z_f$) which
define the range of signals that might be detected 
by present and forthcoming neutrino telescopes: $\fgg^{\rm min}$ and $\fgg = 0.98$ ($z_f = 10$). 
As seen, the total rates range from $4 \times 10^8$ to $8 \times 10^5$ events/yr. If the
jets are beamed into 1 \% of the sky, the effective rate will range between 
$4 \times 10^6$ to $8 \times 10^3$ events/yr.

The dotted lines in Fig.~1 represent the predicted duty cycle of the signal. This is
defined as the ratio of the typical duration of a single burst, $t_{\rm j} \approx 10 (1+z)$~s
(see next Section), to the mean time interval between two successive bursts [$1/R(z)$]. 

The total duty cycle is found to be $> 1$ but, if $z_f \approx 10$, the effective duty cycle might be
significantly reduced by beaming. This means that the diffuse flux
is stationary but might not be continuous in time, being characterized by a sequence of
individual bursts whose average separation depends on the rate.
Integrating over a typical observation time of
$\approx$ 1 yr, the expected number of bursts is $>10^3$ depending
on the assumed values for $\fgg$ and on the beaming factor (see Fig.~\ref{fig00}). 
Therefore, we assume that the average diffuse flux provides a good representation of the signal
averaged over an observation time of 1 yr\footnote{The relative error associated
to this approximation is $\approx 1/\sqrt{N}$ where $N$ is the number of individual
bursts within the observation time}. Signal detection from individual bursts by forthcoming km-scale 
neutrino telescopes will be discussed in Section 6.

\section{The Emission Model}

In this Section we derive the expected neutrino flux contributed by a
single source, \ie a zero-metallicity, rotating, $300 \msun$ star which 
collapses to a VMBH. 

The simulations carried out by Fryer, Woosley \& Heger (2001) show that
the structure of the pre-collapse star is that of a $180 \msun$ He-core
with radius $\approx 6 \times 10^{11}$~cm surrounded by a large H-envelope
which extends up to $\approx 2 \times 10^{14}$~cm. After helium burning,
the core encounters electron-positron pair instability and collapses igniting
explosive oxygen and silicon burning. 
However, a large fraction of the star becomes so hot that photodisintegration
instability occurs. This uses all the energy released by previous nuclear burning
stages and, instead of reversing the implosion, it accelerates the collapse 
and a proto-black hole forms quickly accreting mass from the dense inner part of the core. Centrifugal
forces are able to slow down material at the equator. When the black hole mass 
is $\approx 140 \msun$, a stable accretion disk forms with mass $\approx 30 \msun$ 
and accretion rates of order $1-10 \, \msun \, \mbox{s}^{-1}$. 

It is widely accepted, although not confirmed, that accretion systems of this kind represent one of the leading
models to power the relativistic fireball in GRBs (Woosley 1993). In order to produce
a jet, the energy generated must be transported out by magnetohydrodynamical (MHD) processes. 
This mechanism appears to be the most appropriate to very massive collapsars, 
as other means of energy transport become inefficient as the black hole mass increases 
(Popham \etal 1999). Any quantitative prediction about
the energy and collimation of the magnetic-driven jet is necessarily uncertain because 
of the complex physics involved. The energy released by the jet can be written as
$\epsilon_{\rm disk} \epsilon_{\rm MHD} M_{\rm disk} c^2$ 
where $\epsilon_{\rm disk} \approx 0.11$ is the efficiency of the accretion disk for
typical values of the angular momentum, $\epsilon_{\rm MHD} \approx 0.1$ is the
efficiency of the magnetic-driven explosion and $M_{disk}$ is the mass of the accretion
disk (Fryer, Woosley \& Heger 2001). This yields $\approx 6 \times 10^{53-54}$~erg 
depending on the assumed angular momentum distribution. If beamed into 1\% of the sky, this
jet corresponds to an inferred isotropic energy $E_{\rm iso} \approx 10^{56}$~erg, 
1-2 orders of magnitude larger than for ordinary collapsars. The jet lifetime is limited by disk-fed accretion
and can be crudely estimated to be $t_{\rm j}\approx 10 (1+z)$~s (Fryer,
Woosley \& Heger 2001). Therefore, PopIII GRBs occurring at redshifts $\gsim 10$ are
expected to last longer than their more recent ($\lsim 5$) counterparts. 

Recently, M{\'e}sz{\'a}ros \& Waxman (2001) have proposed a mechanism for the emission of TeV 
neutrinos in collapsars. In the following, we give a brief description of the proposed
model in order to introduce the relevant quantities for the present analysis. 
The interested reader is referred to the original paper for details. 

\vspace{0.5truecm}
\centerline{\it a) Generalities of the model}

In a generic collapsar model, the jet propagates inside the He-core and 
the H-envelope
before emerging. As the jet propagates out into the surrounding circumstellar 
matter, 
shocks are formed which ultimately produce $\gamma$-rays. In their model, 
M{\'e}sz{\'a}ros \& Waxman (2001) show that as the jet is propagating 
through the
dense He-core, it decelerates to subrelativistic velocities. However, 
beyond the
He-core edge at $r \approx 10^{12}$~cm,  the density in the H-envelope drops
down to $\approx 10^{-7} \mbox{g}\,\mbox{cm}^{-3}$ and the jet can accelerate
to relativistic velocity. At the edge of the He-core the jet is capped by a 
termination shock and a reverse shock moves back into the jet, 
accelerating the electrons of the shocked plasma. 
The electrons are expected to lose  all their energy on a very short timescale 
by synchrotron and inverse-Compton emission; the resulting radiation quickly
thermalizes due to the high optical depth to an approximate black body spectrum
whose peak energy is in the X-ray domain.
Inhomogeneities in the jet cause internal shocks at a typical distance 
of $\approx 10^{11}$~cm, which is 
smaller than both the termination and the reverse shock radii
(M{\'e}sz{\'a}ros \& Waxman 2001).
Internal shocks can accelerate protons which interact with the X-ray photons
leading to pion production and consequently to neutrino emission.
We describe this process in detail in the following subsections.

\vspace{0.5truecm}
\centerline{\it b) Target radiation}
Electrons accelerated in the reverse shock, at $r\approx 10^{12}$ cm,
lose  all their energy to radiation which quickly thermalizes  
due to the high optical depth. Therefore the spectrum of the target 
photons is assumed to have a blackbody shape peaking at energy
\be
\epsilon_{\gamma} \simeq 4 \, \Gamma_{\rm h}^{1/2} 
\left(\frac{E_{53}}{r_{12}^2 t_{j,1}}\right)^{1/4} \, \mbox{keV}
\ee
where $E_{53} = E_{\rm iso}/10^{53}$ erg is the isotropic energy
of the explosion, $t_{j,1}=t_{j}/10$ s is the jet lifetime,
$r_{12}=r/10^{12}$ cm is the shock radius, $\Gamma_{\rm h}$ is the
Lorentz factor of the shocked jet plasma; all quantities are
evaluated in the source rest frame. 
 
\vspace{0.5truecm}
\centerline{\it c) Photo-meson interaction}

Internal shocks can accelerate protons to a power 
law distribution which approximates 
$dn_p/d\epsilon_p \propto \epsilon_{p}^{-2},$ where $n_p$ is the number 
of protons with energy $\epsilon_p$ (Waxman 2001).
The maximum proton energy is determined by  equating the acceleration
time, estimated as the Larmor radius divided
by $c$, to the minimum between the dynamical time and the synchrotron
cooling time.
In the source rest frame this is given by,
\be
\label{eq:epmax}
\epsilon_{p,{\rm max}} \approx 9.5 \times 10^{11}
\, \frac{\Gamma_{2.5}^{5/2}\, \delta t_{-2}^{1/2}\, t_{j,1}^{1/4}}
{(E_{53}\xi_{\rm{B},-2})^{1/4}}\,{\rm GeV},
\ee
where $\Gamma_{j}=10^{2.5}\Gamma_{2.5}$ is the typical Lorentz factor 
of the jet, $\delta t = 10^{-2} \delta t_{-2}$ s 
is the variability timescale of the injected relativistic outflow
and $\xi_{\rm B}=10^{-2}\xi_{{\rm B},2}$ is the magnetic field equipartition
fraction.
As they approach the reverse shock, high energy protons may produce 
both charged and neutral pions, with roughly equal probabilities,
through the interaction with photons of energy 
$\epsilon_\gamma$,
\be
p +\gamma \rightarrow \pi^0 +p 
\qquad \qquad \,\,\,\, p +\gamma \rightarrow \pi^+ +n.
\ee
In order for this reaction to take place, the proton  energy must exceed
the threshold energy for $\Delta$-resonance with thermal photons, \ie
\be
\label{eq:delta}  
\epsilon_p \gsim \epsilon_{p,\Delta} = 0.3 \, \Gamma_{\rm h}^2 \, \epsilon_\gamma^{-1}\,\, {\rm GeV}.
\ee
These protons will lose  all their energy to pions through multiple collisions 
at a typical distance from the termination shock where the photon density is 
such that 
the photo-meson optical depth $\tau_{p,\gamma} \gg 1.$
Charged pions will then decay into muons finally leading to neutrino
production:
\be
\pi^+ \qquad \rightarrow \qquad \mu^+\ +\ \nu_{\mu} \qquad \rightarrow \qquad 
e^+ \ +\ \nu_e\ +\ \bar{\nu}_\mu+\ \nu_{\mu}.
\ee
In a single collision, a  pion is created with an  average energy 
which is $20\%$ of the proton energy. 
This energy is roughly evenly distributed between the
final $\pi^+$ decay products, yielding
a neutrino energy  $\epsilon_{\nu} = 0.05 \epsilon_p$.

\vspace{0.5truecm}
\centerline{ \it d) Neutrino energy spectrum}

The neutrino energy spectrum emitted by each source can be expressed as 
\be
J_{\nu} = \epsilon_{\nu}^2 \frac{dn_{\nu}}{d\epsilon_{\nu}},
\ee
where the neutrino energy ranges between a minimum value,
$\epsilon_{\nu,\rm{min}} = 0.05 \epsilon_{p,\Delta}$,
and a maximum value $\epsilon_{\nu,\rm{max}} = 0.05 \epsilon_{p,{\rm max}}$
and $n_{\nu}$ is the number of neutrinos with energy $\epsilon_{\nu}$.

Inverse Compton losses might be relevant for high energy pions and
muons.
From the results given in M{\'e}sz{\'a}ros \& Waxman (2001),
we derive that while pions are expected to decay prior to 
significant energy loss, muons, whose lifetime is 
$\approx 100$ times longer, may lose  significant fraction of 
their energy before decaying. 
To be conservative we consider only those neutrinos which are
directly produced by $\pi^+$ decay; these carry $\sim 1/4$ of the pion
energy. 
As a consequence, the total energy emitted in neutrinos is given by
\be
E_{\nu,\rm{tot}}\simeq \frac{\xi_p }{8}
\frac{\ln (\epsilon_{\nu,\rm{max}}/\epsilon_{\nu,\rm{min}})}
{\ln (\epsilon_{p,\rm{max}}/\epsilon_{p,\rm{min}})}\, E_{\rm iso},
\label{eq:enutot}
\ee
where $\epsilon_{p,{\rm min}} = \Gamma_{\rm h} m_p c^2$, 
$\xi_p \approx 0.4$ is the fraction of the total energy
in accelerated protons, assuming equipartition.

Synchrotron losses are unimportant for protons within the present energy
range but might become relevant for pions with energy greater than 
a threshold value above which pions lifetime exceeds the characteristic
time for energy loss due to synchrotron emission 
(Waxman \& Bahcall 1997, 1999; Rachen \& M{\'e}sz{\'a}ros 1998). The corresponding
neutrino energy is given by,
\be
\epsilon_{\nu,s} = 10^8 
\left[\frac{\Gamma_h\,\Gamma_{2.5}^7\, \delta t^2_{-2}\, t_{j,1}}
{E_{53}\xi_{\rm{B},-2}}\right]^{1/2}\,{\rm GeV}.
\ee

Since $\tau_{p,\gamma} \gg 1$, at all energies $\epsilon_{\nu} \leq \epsilon_{\nu,s}$, 
the $\nu$ energy spectrum is simply proportional to that of 
the protons, \ie $dn_{\nu}/d\epsilon_{\nu} \propto \epsilon_{\nu}^{-2}$ and we can write,

\[
J_{\nu} = \left\{ \begin{array}{l}
{\cal N} E_{\nu,\rm{tot}} \qquad \qquad  \qquad \,\, \, {\rm if}  \qquad \epsilon_{\nu, \rm min}\le \epsilon_{\nu} \leq \epsilon_{\nu,s} \\
{\cal N} E_{\nu,\rm{tot}}\left[\epsilon_{\nu}/\epsilon_{\nu,s}\right]^{-2} \qquad {\rm if} \qquad \,\,\,\, \epsilon_{\nu,s} < \epsilon_{\nu} \leq \epsilon_{\nu,\rm max}; 
\label{eq:jmu} \end{array} \right. \]

$E_{\nu,\rm{tot}}$ is the total energy emitted in neutrinos with 
$\epsilon_{\nu,\rm{min}} \le \epsilon_{\nu} \le \epsilon_{\nu,\rm{max}}$ and \
${\cal N}^{-1} = \ln (\epsilon_{\nu,s}/\epsilon_{\nu,\rm{min}})+1/2[1-(\epsilon_{\nu,s}/\epsilon_{\nu,\rm{max}})^2]$ is a normalization factor.
At neutrino energies $\epsilon_{\nu}>\epsilon_{\nu,s}$, the probability that a 
pion would decay before losing its energy decreases as 
$\approx \left[\epsilon_{\nu}/\epsilon_{\nu,s}\right]^{-2}$ 
and the spectrum is no longer flat.

The average energy flux, $F_{\nu}(E)$ (units GeV cm$^{-2}$), observed at a given energy $E$ and 
emitted by a source at redshift $z$ can be expressed 
as,
\be
F_{\nu}(E)  = 
\int \frac{d\Omega}{4 \pi} \frac{d J^{\rm obs}_{\nu}[E(1+z)]}{d\Sigma} = (1+z)^{-1}\frac{J_{\nu}[E(1+z)]}{4 \pi r^2(z)}  
\label{eq:single}
\ee
for $\epsilon_{\nu,\rm{min}} \le E(1+z) \le \epsilon_{\nu,\rm{max}}$,
and surface element $d\Sigma = r^2(z) d\Omega$, where $r(z)$ is the comoving 
distance to the source, $E (1+z)$ is the emission 
energy, $J_{\nu}[E (1+z)]/
(1+z)$ is the redshifted energy spectrum and beaming effects are properly 
taken into account.

\section{Neutrino Diffuse Flux from PopIII GRBs}

In this Section we estimate the diffuse flux of neutrinos emitted by PopIII GRBs 
through the mechanism described above. The integrated signal observed
today at energy $E$ is 

\be
E^2 \frac{dN}{dE} \,\, [{\rm GeV \, cm}^{-2}\, {\rm s}^{-1}\, {\rm sr}^{-1}] = \frac{\Omega_j}{4 \pi} \int_{z_f} 
dR(z) F_{\nu}(E),
\ee
where $dR(z)$ is the differential rate of all PopIII GRBs, given by 
eq.~\ref{eq:rate}, $\Omega_j = \Delta \Omega_j / 4 \pi = 0.01$ 
is the fraction of sources which are beamed towards the Earth and $F_{\nu}(E)$ is the 
average flux emitted by a single source at energy $E(1+z)$ (see eq.~\ref{eq:single}). 
The lower limit $z_f$ is the transition redshift from a top-heavy to a normal IMF marking
the end of the epoch of VMBH formation; the value of $z_f$  depends on $\fgg$ (see eq.~\ref{eq:IMF}). 
In computing the diffuse flux we are making the implicit assumption that all
sources contribute the same neutrino flux to the overall signal, \ie all sources have the same progenitor 
($300 \msun$) and release the same amount of energy $E_{\rm iso}$.

We show in Fig.~\ref{fig01} the diffuse flux obtained assuming two possible values for
the equivalent isotropic energy $E_{\rm iso} = 10^{54}, 10^{56}$~erg and 
$\fgg^{\rm min}$ that corresponds to the maximum source emission
rates plotted in Fig.~\ref{fig00}. The experimental upper limits for present (AMANDA-B10, hereafter 
always indicated as AMANDA-I) and forthcoming
(AMANDA-II, IceCube) neutrino telescopes are also shown, together with the diffuse flux from atmospheric
neutrinos and a number of other astrophysical sources. The effect of a smaller value for $E_{\rm iso}$ (smaller
explosion energy or larger beaming angle) is to decrease the overall amplitude of the diffuse emission flux
but at the same time to extend the range of observed energies where the spectrum would appear as flat.

As can be inferred from Fig.~\ref{fig01}, 
because of the larger explosion energy, the integrated signal from PopIII GRBs is higher 
than that generated from ``ordinary'' GRBs, at least for a significant range of values 
for $\fgg$ (or, alternatively $z_f$). The maximum flux predicted by our model is above the upper limit recently established
by the AMANDA-I detector (Andr\'es \etal 2001). This finding, together with the anticipated sensitivities of AMANDA-II
(Barwick 2001) and IceCube (Spiering 2001) shows how neutrino telescopes might enable to derive important constraints 
on the characteristic energies released by PopIII GRBs as well as on the IMF of the first stars. 
The relevant quantities that determine the amplitude (and, to a less significant extent, the energy range) of
the diffuse flux are represented by the assumed explosion energy and by the PopIII GRBs IMF. 
As discussed in the previous Section, $E_{\rm iso}$ depends on the rather uncertain physics of 
MHD processes through the quantities $\epsilon_{\rm MHD}$, 
$\epsilon_{\rm disk}$ and $M_{\rm disk}$ as well as on the assumed degree of beaming. 
Therefore, it is equally important to explore a wide range
of values for the explosion energy, such as $10^{52-56}$~erg. The aim of the 
analysis presented in the next Section is to 
investigate to what extent present and forthcoming neutrino telescopes might be able to test our scenario.
In particular, if the current scenario is correct, important considerations regarding 
$E_{\rm iso}$ and $\fgg$ might be derived even from a non-detection, 
\ie if the signal is below the expected upper limits. This will enable to reject
a range of models which so far appear to be untestable by any other observational facility.

\section{Experimental Predictions}

The diffuse emission from unresolved PopIII GRBs (as well as from other astrophysical sources) 
should have an energy spectrum that is much harder than the atmospheric neutrino spectrum 
(see Fig.~\ref{fig01}). Therefore, a search for a diffuse flux can be undertaken by looking
for an excess of events at large energies (see \eg Andr{\'e}s \etal 2001). Unfortunately,
the energy resolution of a neutrino telescope is rather poor. Indeed, the energy of the event
is roughly estimated counting the number of optical modules triggered by Cerenkov 
radiation from muons that are produced in neutrino-nucleon interactions in the ice 
surrounding the detector or in the bedrock below (Andr{\'e}s \etal 2000). Thus, in the
following, we consider two possible energy intervals: [$10^4 - 10^5$] GeV and 
[$10^5 - 10^6$] GeV that we indicate as low- (L) and 
high- (H) energy bands, respectively. Within these energy bands, the
diffuse emission from PopIII GRBs is higher than the predicted experimental sensitivities, at least
for a range of model parameters (see Fig.~\ref{fig02}). The atmospheric neutrino 
emission in the L-band is the primary source of contamination as 
the contribution from other astrophysical sources is well below the 
experimental sensitivity (see Fig.~\ref{fig01}).
 
In Fig.~\ref{fig02} we show the differential number of neutrinos, $dN/dE$, predicted by our
model assuming the same model parameters as in Fig.~\ref{fig01}, 
\ie $E_{\rm iso}=10^{56}$~erg and $\fgg^{\rm min}$. 
The same quantity corresponding to the atmospheric neutrino diffuse flux and to the predicted upper limits 
for AMANDA-I (upper panel), AMANDA-II (medium panel) and IceCube (lower panel) are shown both in the L- and the H-bands. The shaded areas indicate the
range of models corresponding to the same $E_{\rm iso}$ but to different values for $\fgg$
that might lead to signal detection or, equivalently, that might be rejected by a non-detection. 
As can be inferred from the Figure, the upper limit set by AMANDA-I (Andr{\'e}s \etal 2001) can
be used to reject a region of model parameters: if $E_{\rm iso} = 10^{56}~{\rm erg}$,
than $\fgg$ must be $>0.27$ and the transition to a normal IMF must have occurred at redshifts $z_f > 9.2$. 

The building of km-scale detectors will significantly improve the sensitivities
in both bands, even though the L-band appears to be largely dominated by the atmospheric neutrino diffuse flux. 
Thus, AMANDA-II and IceCube are expected to show equivalent performances in this channel.
Conversely, because of the softer energy spectrum, the atmospheric neutrino signal drops
in the H-band and IceCube is expected to significantly improve the experimental constraints that might be 
achieved by AMANDA-II, as can be seen by comparing the extension of the shaded regions in 
Fig.~\ref{fig02}.

The effect of variations in the values of  $E_{\rm iso}$ and $\fgg$ can also be explored.
To make a systematic analysis of the parameter space, for each model 
[identified by a point in the ($E_{\rm iso}$, $\fgg$) plane] we compute the quantity $R_{\rm L}$ 
($R_{\rm H}$) defined as the total number of neutrinos in the L-band (H-band) normalized to the 
largest between the number of atmospheric neutrinos and the experimental upper limits for 
AMANDA-I, AMANDA-II and IceCube. 

Figs.~\ref{fig03} and ~\ref{fig04} show maps of $R_{\rm L}$ (upper panels) and $R_{\rm H}$ (lower panels) 
for AMANDA-II and IceCube. The solid lines mark the limits of the regions (right to the curves) 
where these detectors would be able to detect a count excess. The dashed lines show the current limit
set by AMANDA-I, which corresponds to explosion energies $E_{\rm iso}$ in the range 
$[6.4 \times 10^{54}-10^{56}]$~erg and to $\fgg \le [8 \times 10^{-3}-0.27]$ depending on $E_{\rm iso}$. 

Thus, according to the present observational constraints, if PopIII GRBs had occurred
with energies in the above range, than the fraction of \SNgg to VMBHs must have been higher than the 
above limit (\ie the transition from a top-heavy to a normal IMF must have occurred at redshifts 
$z_f \ge [5.4 - 9.2]$ depending on $E_{\rm iso}$) in order for the overall signal to be below the experimental upper limit.
 
In the L-band, AMANDA-II and IceCube would be able to extend the region excluded by AMANDA-I down to
energies $3.6 \times 10^{54}$~erg and up to $\fgg \le 0.43$ (transition redshifts $z_f \approx 9.75$).
Moreover, the enhanced sensitivities of km-scale detectors open the possibility to test our scenario
in almost all range of plausible explosion energies ($E_{\rm iso}\ge 7.6 \times 10^{52}$~erg) and
to derive fundamental insights on the IMF of the first stars, with $\fgg$ ranging from
$8 \times 10^{-3}$ up to 0.98.

Thus, unless MHD efficiencies are so low to significantly reduce the energy released by PopIII GRBs,
the anticipated sensitivities of AMANDA-II and IceCube will very likely lead to a direct detection of  
PopIII GRBs in at least one of the two energy bands. The above range of values on $\fgg$ corresponds to
two opposite scenarios for the evolution of the first stars: if $\fgg \approx 8 \times 10^{-3}$ than
approximately 1 \SNgg \, forms every 120 VMBHs (hence PopIII GRBs). 
This means that a very small amount of mass has been processed into metals and that star formation 
proceeds according to a top-heavy IMF down to redshifts $z_f \approx 5.4$. Since the bulk of the mass 
went in PopIII progenitors with masses $> 260 \msun$, we expect a large number of VMBHs remnants
to be present today. The resulting critical density in VMBHs has been estimated to be high enough
to account for the entire baryonic dark matter halos of present-day galaxies (see Schneider \etal
2001 for a thorough discussion of these issues). Conversely, if $\fgg \approx 0.98$, than approximately
1 VMBH forms every 75 \SNgg. In this case, a large fraction of mass in PopIII stars has been processed
into metals and transition to a normal IMF occurs at much higher redshift, at $z_f \approx 10$. However,
the critical density in VMBHs formed at these early stages is still comparable to that of 
super massive black holes observed in the nuclei of present-day galaxies (Schneider \etal 2001). 
 
\section{Implications for PopIII Gamma Ray Bursts}

At present it is not clear if the relativistic jets can break-out of the stellar 
envelope. In PopIII GRBs progenitors, jet driving depends mainly upon MHD 
processes. For this reason, such jets 
tend to be relatively cold, in the sense that their thermal energy is small
compared to their kinetic energy (MacFadyen, Woosley \& Heger 2001) and
hence they are efficiently pressure-collimated by the He-core through 
which they propagate. 
However, the internal structure of PopIII stars is different from that of ordinary collapsars,
being characterized by larger He-core densities and more
extended H-envelopes. Under these conditions, it is likely that jet stalling 
might occur before break-out if the crossing
time is longer than the jet lifetime. Since the jet becomes relativistic near the outer
radius of the He-core, a more extended H-envelope does not affect its crossing time, which
is substantially determined by the fly time through the He-core (M{\'esz{\'a}ros \& Rees 2001).  
However, the effect of rotation in PopIII GRBs is to create a funnel along the rotation
axis within which the density becomes substantially lower than the equatorial one. Therefore,
at least in some cases, this might lead to a break-out.
If this is the case, GRBs are a necessary endproduct of such process; 
if instead the jet is chocked the neutrino signal will nevertheless be emitted.  
In the previous Sections we have already analyzed in detail the latter
scenario. Here we briefly analyze the implications descending from a
successful jet break-out, leading to a population of very high redshift GRBs. 
What are the distinctive signatures that would enable us to pin-point PopIII GRBs
among observed ones? 
    
Probably the most obvious such features are the following: {\it i)}
depending on the properties of the accretion
disk and on the uncertain interaction of the jet with surrounding matter, 
the PopIII GRBs should probably on average last longer, due to cosmological time dilation; 
{\it ii)} the peak of emission, 
that in the rest frame is in $\gamma$-rays, would be shifted into X-rays; 
{\it iii)} their optical afterglow should be heavily absorbed by the intervening
intergalactic medium (particularly if they occur before reionization); {\it iv)} iron lines
should be very weak or totally absent, because nucleosynthetic products of progenitor
stars are swallowed by the VMBH and because the surrounding gas is presumably of 
primordial composition.
Have events with similar characteristics already been observed?

Intriguingly, BeppoSAX has revealed the existence of a new class of events, 
the so-called X-ray flashes or Fast X-ray Transient (FXTs), which show the 
bulk of emission in X-rays (Heise \etal 2001). 
They are bright X-ray sources, with peak energies  
in the $2-10$~keV band and durations between $10-200$~s, which are not triggered and not detected in the 
$\gamma$-ray range $40-700$~keV. A subsample of 10 FXTs have been observed in the lowest energy channels
of the BATSE instrument (Kippen \etal 2001). This has enabled to directly compare the gamma-ray
properties of FXTs to those of ``ordinary'' GRBs. The preliminary result is that FXTs appear to
be consistent with an extrapolation of GRB behaviour, showing softer spectra (based on their fluence hardness
relation) and duration all in excess of $10$~s. While the limited sample does not allow to    
draw statistically solid conclusions, none of the observational evidences is in apparent conflict 
with their nature being PopIII GRBs. Moreover, peak energies in the observed range ($2-10$~keV) are
naturally predicted in PopIII GRBs not only as a consequence of higher emission redshifts. Indeed,
gamma-ray emission will occur at larger radius because of the more extended H-envelope, leading to
a lower value for the emitted peak energy ($\approx 200$~keV, Guetta, Spada \& Waxman 2001).
In addition, requirement {\it iii) } above
is very likely fulfilled. In fact, a significant number (about 40\% of GRBs for which 
fast follow-up observations were carried out) of GRBs do not show an optical counterpart; 
for this reason they are dubbed as GHOST (GRB Hiding Optical Source Transient). 
It is quite interesting to note that all FXTs fall into the GHOST category\footnote{As we are writing, an X-ray rich GRB (GRB011211) has been detected by BeppoSax WFC1. Follow-up observations 
have identified the optical afterglow and have measured a redshift $z=2.1$ 
(Piro, private comunication). However, a detailed analysis of the BeppoSAX GRBM data shows 
a very shallow long event in the 40-700 keV band with total duration of about 270 s,
similar to that measured in X-rays (2-28 keV) by WFC1. Given the redshift,
the (isotropic) $\gamma$-ray energy is $6.3 \times 10^{52}$~erg 
(Frontera \etal GCN 1215). This preliminary analysis shows that the X/$\gamma$ 
fluence ratio is similar to that found in a number of ordinary GRBs detected by BeppoSAX.
The distinguishing features of GRB011211 are its long duration (the
longest event localized with BeppoSAX) and its faintness both in X- and
$\gamma$-rays. Therefore, the analysis leads to conclude that 
this event cannot be classified as a FXT.}

The above scenario cannot yet be considered more than a promising possibility 
and there are alternative explanations. For example, the extension of GRBs to the low peak energy regime
may be a natural consequence of the fireball model.
As pointed out by Guetta, Spada \& Waxman (2001), FXTs could be also produced
by relativistic winds where the minimum Lorentz factor is smaller
than that required to produce ``normal" GRBs. The region of parameter space 
leading to FXTs in their model is rather small; in addition, they do not make
any specific prediction on the burst duration. In this framework,   
the failed optical detection can be caused by dust extinction within the host
system, although this explanation is still debated (Lazzati,
Covino \& Ghisellini 2001; Djorgovski \etal 2001, Ghisellini 2001). 
Prompt infrared observations would be able to tackle this issue.

The number of undetected FXTs is estimated to be very large. 
The untriggered BATSE catalogue suggests an all-sky rate of $\approx 400 {\rm ~yr}^{-1}$ 
(Stern \etal 2000). If we maintain the above speculation that FXTs are indeed successful 
(\ie non chocked) PopIII GRBs, and if a neutrino precursor from a single event could be
measured in coincidence with an X-ray flash, this would provide a direct probe of the nature of 
X-ray flashes and on the relative number of chocked and successful GRBs.
An estimate of the likelihood of such an event can be given as follows: the average 
number of upward muon detections per unit area in a neutrino telescope during an individual burst
is
\[
N_{\mu} = P_{\nu\mu} \frac{F_{\nu}(E)}{E},
\]
where $P_{\nu\mu}=1.3 \times 10^{-6} (E/{\rm TeV})^{\beta}$ ($\beta=2$ if $E<1$ TeV and $\beta=1$ if $E>1$ TeV)
is the probability that a muon neutrino will produce an upward moving muon in a detector
(Gaisser, Halzen \& Stanev 1995) and $F_{\nu}(E)$ is the average neutrino flux emitted by a single source
as given by eq.~\ref{eq:single}. For a neutrino energy spectrum corresponding to 
$E_{\rm iso} = 10^{56}$~erg, $N_{\mu}$ is constant with energy in the L-band (see Fig.~\ref{fig01}).  
Assuming a ${\rm km}$-scale detector (such as IceCube) and a source energy, the number of muon detections per burst 
is $N_{\mu} = 0.03 - 0.003 $ for emission redshift 
$5.4 \le z \le 30$. 
This signal is significantly higher than both the atmospheric
neutrino background and the cumulative PopIII GRBs signal at comparable energies. Since the typical
angular resolution of planned neutrino telescopes is $\theta \approx 1^{\circ}$, a diffuse flux will produce
a negligible average number of muons around the direction of arrival from a point source within the duration
of the burst. 
Assuming the above mentioned rate of $400 {\rm ~yr}^{-1}$ for FXTs, one could
expect a few neutrino bursts/yr with $N_{\mu} > 1$. 
This result is obtained assuming that all FXTs are successful PopIII GRBs beamed towards the Earth, which for
a beaming factor of 1\% and values for $\fgg$ in the detectable region, implies a ratio of successful to chocked GRBs of less than 5\%.

\section{Summary of the Results}

The main results of this paper can be summarized as follows: 

$\bullet$ The total rate of PopIII GRBs produced by massive, first stars can 
be as high as $4 \times 10^6$~yr$^{-1}$ for the most favourable parameter choice and 
if the jets are beamed into 1\% of the sky; 
for some model parameters, the event duty cycle is found to be smaller than unity implying that the 
diffuse neutrino signal might be characterized by a sequence of individual bursts.

$\bullet$ High energy neutrinos from PopIII GRBs could dominate (above
other plausible neutrino sources) the flux  
in the L [$10^4-10^5$] GeV and H [$10^5 - 10^6$] GeV energy bands which can be explored by
present (AMANDA-I) and forthcoming (AMANDA-II, IceCube) neutrino telescopes (Fig.~\ref{fig02}).

$\bullet$ Constraints on $\fgg$ (\ie on the mass fraction of PopIII stars which ends up as
SN$_{\gamma\gamma}$) from these experiments will also provide
a tremendous insight into the IMF of the first stars, and in particular,
on the redshift $z_f$, which marks the metallicity-driven transition from a top-heavy
IMF to a the current, standard (\ie Salpeter-like) one.

$\bullet$ For explosion energies in the range $E_{\rm iso} = 6.4 \times 10^{54} -10^{56}$~erg, the
current upper limit established by AMANDA-I in the L- and H-bands can be used to reject a region
of the model parameter space, namely $\fgg \leq [8 \times 10^{-3}-0.27]$ corresponding to 
transition redshifts $z_f \leq [5.4-9.2]$. Thus, either PopIII GRBs are characterized by smaller
explosion energies or massive primordial star formation occurred only at very high redshifts
and metals released by \SNgg ~~enriched the IGM to the critical level $<Z> = 10^{-4} Z_{\odot}$ 
before redshift 9.2.

$\bullet$ The enhanced sensitivities of forthcoming experiments such as AMANDA-II and, in particular,
km-scale detectors such as IceCube, might be able to discriminate between the two above possibilities: 
observations in the H-band
will explore a wide range of explosion energies, $E_{\rm iso}=7.6 \times 10^{52}-10^{56}$~erg,
values for $\fgg$ in the range $8 \times 10^{-3} \le \fgg \le 0.98$ and transition
redshifts $5.4 \le z_f \le 10$.

$\bullet$ For a subset of model parameters, neutrino emission from individual bursts might be detected 
with km-scale detectors in the L-band over roughly a year of observations. 
To be statistically significant, single burst detection should occurr in coincidence with
observations of the GRB itself, if the jet is not chocked and can successfully propagate out of 
the stellar envelope.

$\bullet$ We have speculated that PopIII GRBs, if they are not chocked, could be associated
with a new class of events, the Fast X-ray Transient (FXTs), which 
are bright X-ray sources, with peak energies
in the $2-10$~keV band and durations between $10-200$~s, which are not 
triggered and not detected in the
$\gamma$-ray range $40-700$~keV. An additional hint that they could be high 
redshift sources comes from the fact that all FXTs observed so far fall into the GHOST  
(GRB Hiding Optical Source Transient) category, an occurrence highly suggestive
on the absorbing effects of the predominantly neutral intergalactic medium 
present at epochs prior to cosmic reionization.

\begin{acknowledgments}

We are grateful to B. Ciardi, C. di Stefano, K. Mannheim, G. Riccobene, C. Spiering
for stimulating discussions and useful information. We particularly acknowledge 
L. Piro and E. Waxman for useful conversations and ideas which greatly improved this work.
This work was partially supported (RS) by the Italian CNAA (Project 16/A).    

\end{acknowledgments}

\begin{figure}
\centerline{\psfig{figure=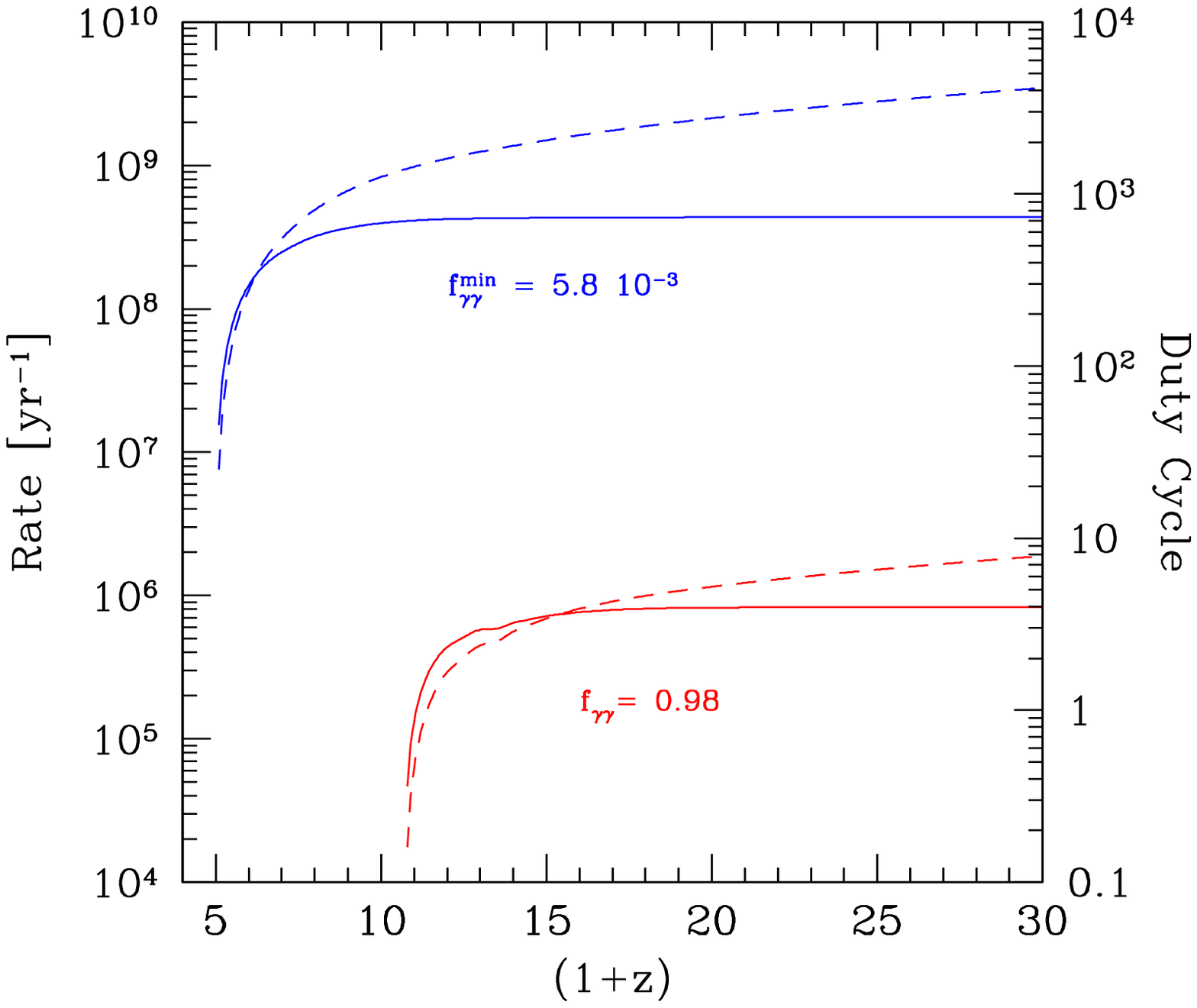,height=17cm}}
\caption{\footnotesize   
Predicted total rate of Pop III GRBs as a function
of redshift is shown (solid lines) for two different values of
$\fgg$ (hence $z_f$, see text). For the same 
values, we plot (dashed lines) the duty cycle of the signal
assuming the duration of each burst to be $10 (1+z)$~s.}
\label{fig00}
\end{figure}
\begin{figure}
\centerline{\psfig{figure=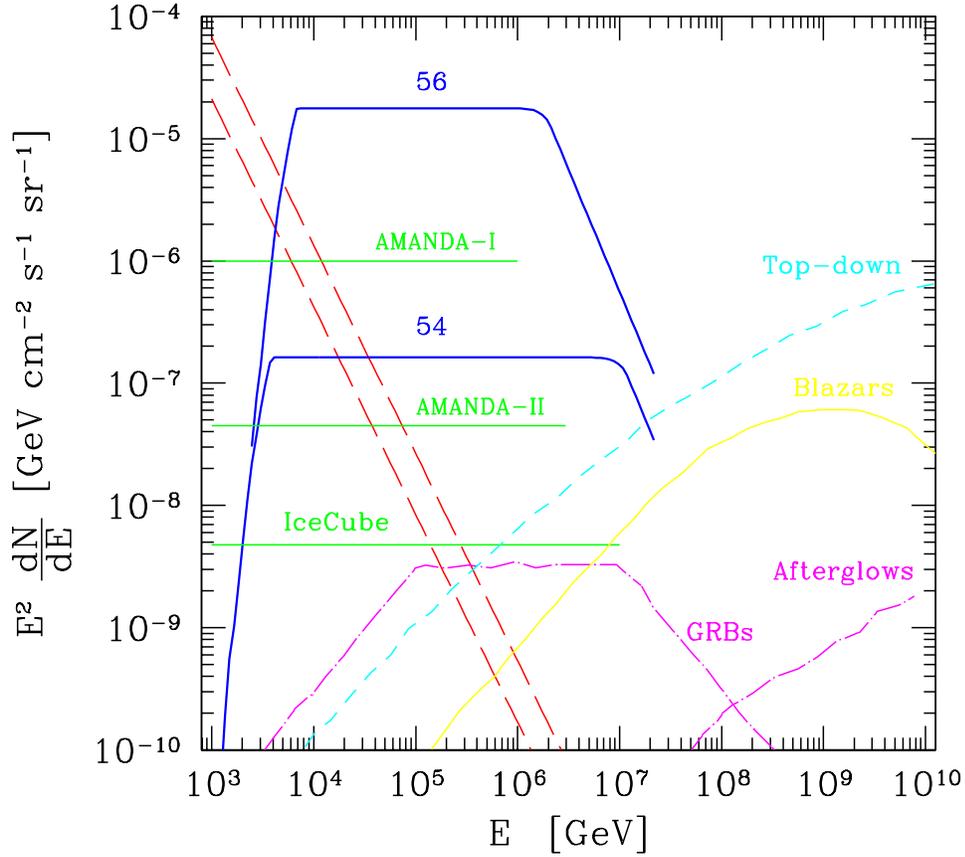,height=17cm}}
\caption{\footnotesize The diffuse neutrino flux from different
sources is compared to the sensitivities of present and forthcoming 
telescopes. The diffuse emission from Pop III GRBs is computed 
assuming two possibile values for the explosion energy, 
$E_{\rm iso} = 10^{54}, 10^{56}$~erg (solid curves labeled 54 and 56 respectively)  
and  $\fgg^{\rm min}$ that corresponds 
to the maximum source rate (see Fig.~\ref{fig00}). 
The long-dashed lines represent the diffuse emission
of atmospheric neutrinos in the horizontal (upper boundary) and vertical 
(lower boundary) direction (Lipari 1993). The other models correspond to
the diffuse emission from ``ordinary'' GRBs and GRBs afterglows (Waxman \& Bahcall 1997),
to a model for photo-meson interaction in blazar jets producing ultra-high energy
cosmic rays through neutron escape  and to 
decaying gauge bosons created at topological defects (see Learned \& Mannheim 2000
and references therein).} 
\label{fig01}
\end{figure}
\begin{figure}
\centerline{\psfig{figure=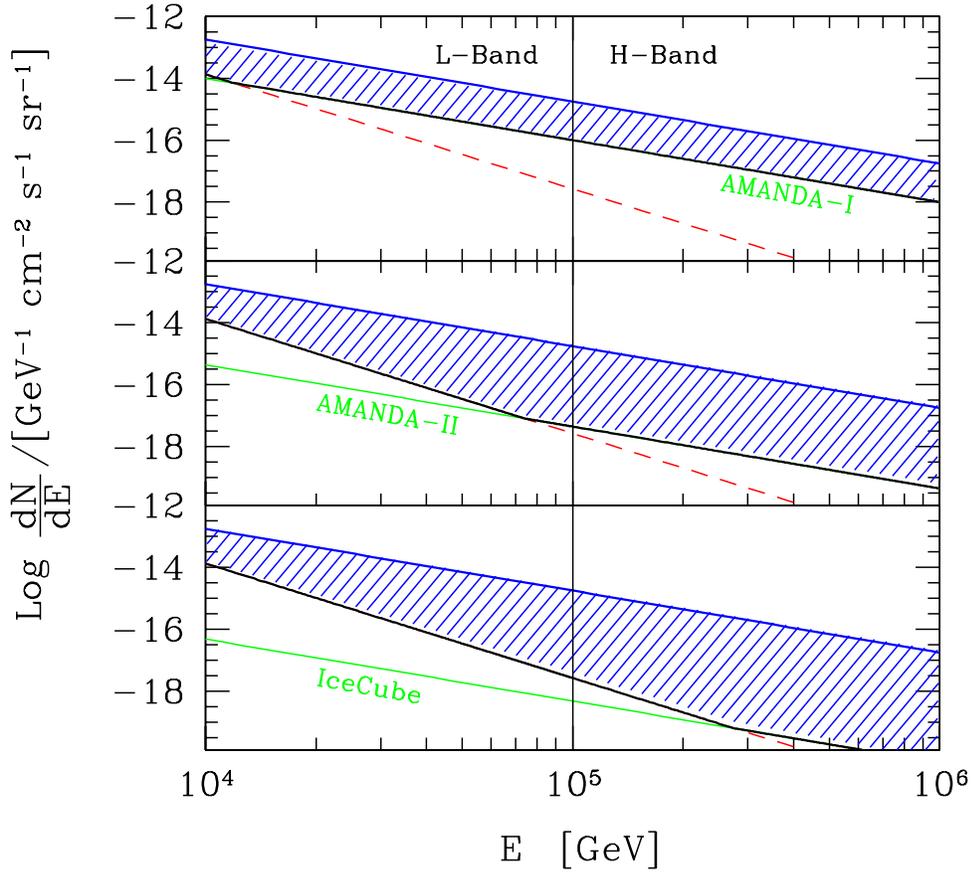,height=17cm}}
\caption{\footnotesize The number of neutrinos per energy
interval as a function of the observed energy predicted by
our reference model ($E_{\rm iso} = 10^{56}$~erg) assuming $\fgg^{\rm min}$ 
(maximum source rate). The signal
is compared to the atmospheric neutrino diffuse emission (dashed line, corresponding
to the upper bound in Fig.~\ref{fig01}) and to the sensitivities of AMANDA-I (upper panel), 
AMANDA-II (medium panel) and IceCube (lower panel). 
The shaded areas identify the regions of parameter space
(corresponding to the same $E_{\rm iso}$ but to different $\fgg$) that would appear
as an excess of events in the low and high-energy bands.}
\label{fig02}
\end{figure}
\begin{figure}
\centerline{\psfig{figure=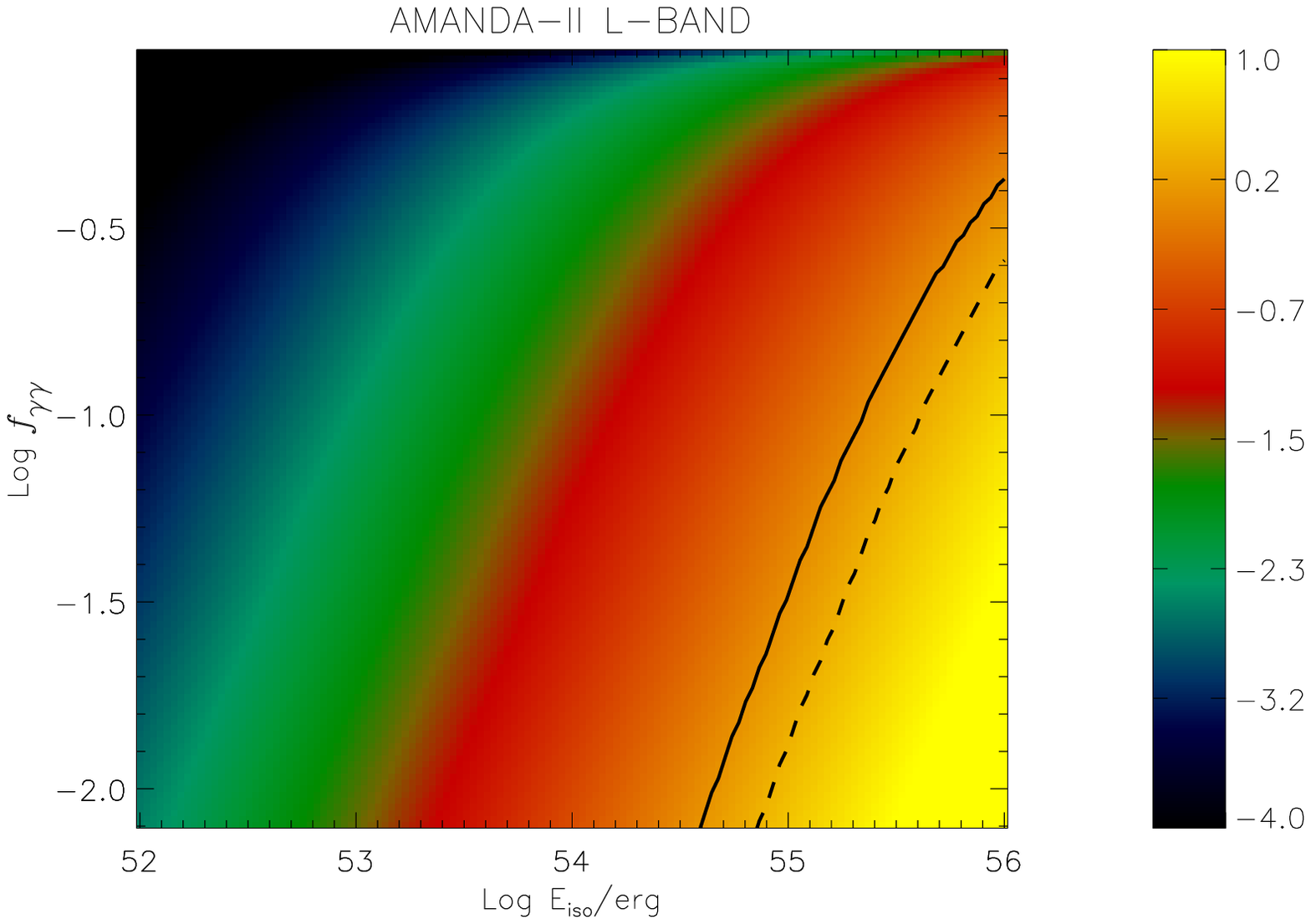,height=7.5cm}}
\centerline{\psfig{figure=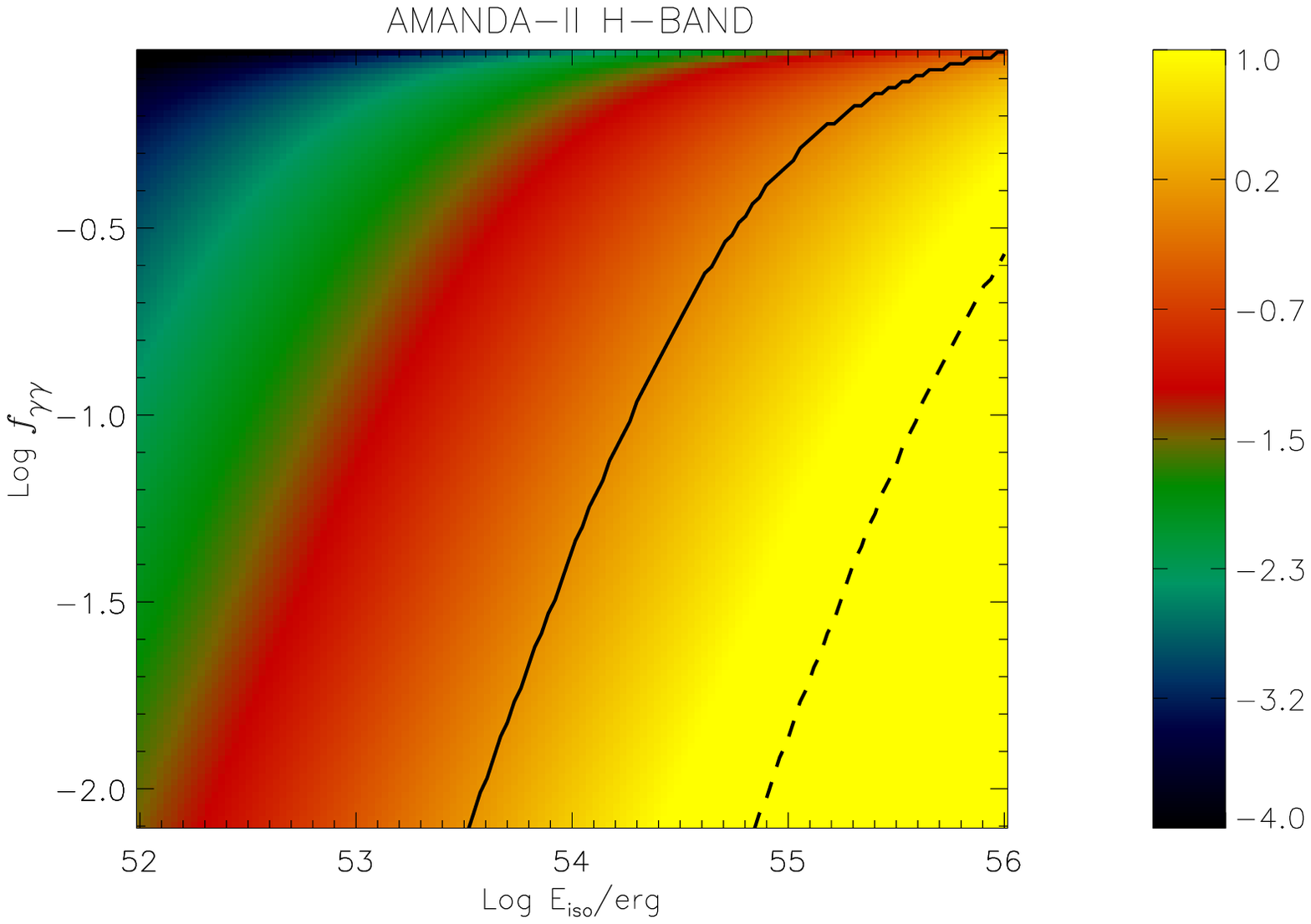,height=7.5cm}}
\caption{\footnotesize {\it Upper panel}: values of $R_{\rm L}$  
i.e. the ratio of predicted
number of neutrinos in the low energy band (L: $10^4-10^5$~GeV) normalized 
to the predicted AMANDA-II sensitivity in the same band, in the  
($E_{\rm iso}, \fgg$) plane.  The color bar shows $R_{\rm L}$ in Log scale. 
The solid line identifies the region (right to the curve) where 
$R_{\rm L} > 1$; the region rightmost of the dashed line is excluded by the AMANDA-I current limit.
{\it Lower panel}: same as above but for $R_{\rm H}$, {\it i.e.} the ratio of
predicted number of neutrinos in the high energy band (H: $10^5-10^6$~Gev) to 
AMANDA-II sensitivity in the same band.}
\label{fig03}
\end{figure}
\begin{figure}
\centerline{\psfig{figure=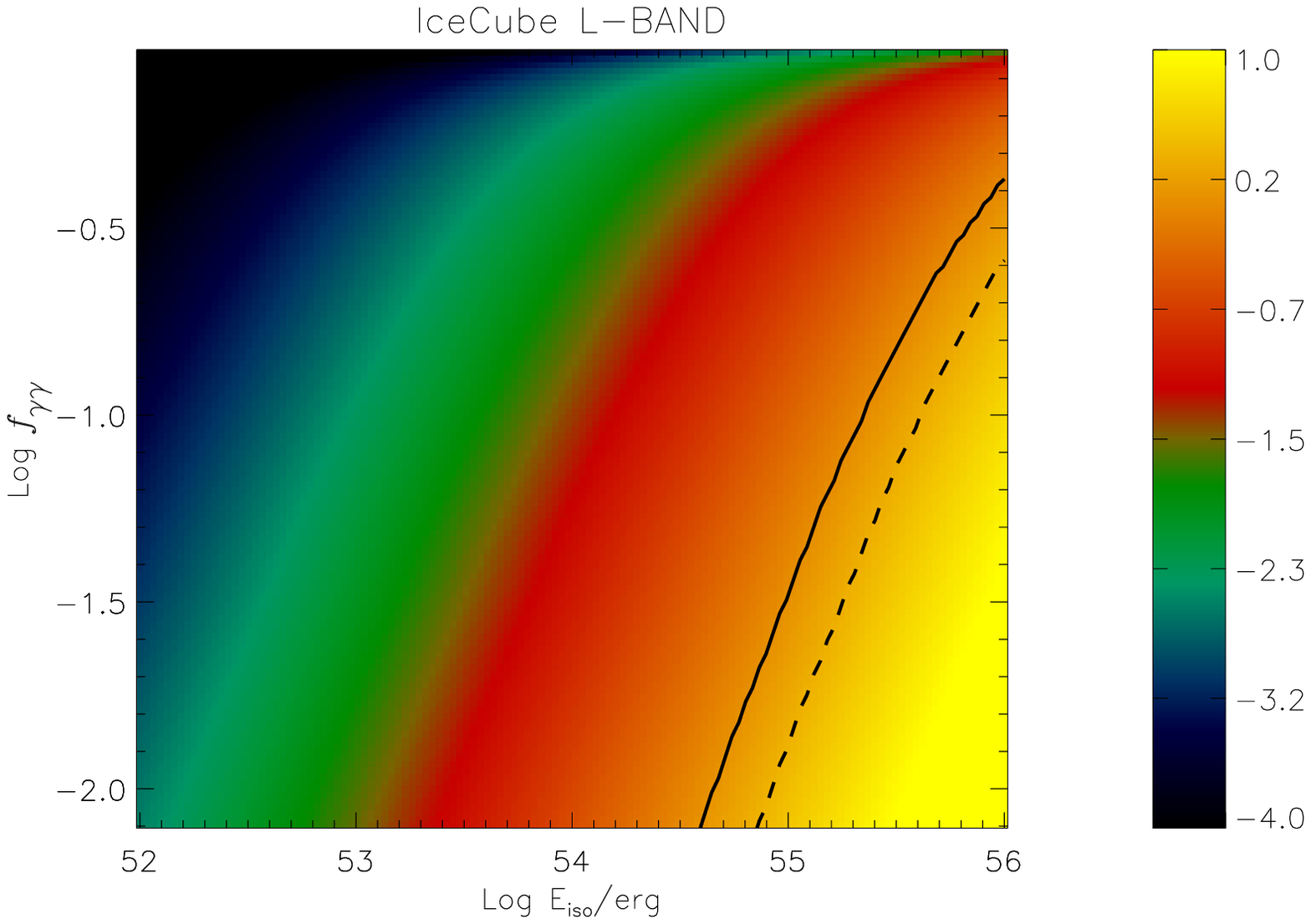,height=7.5cm}}
\centerline{\psfig{figure=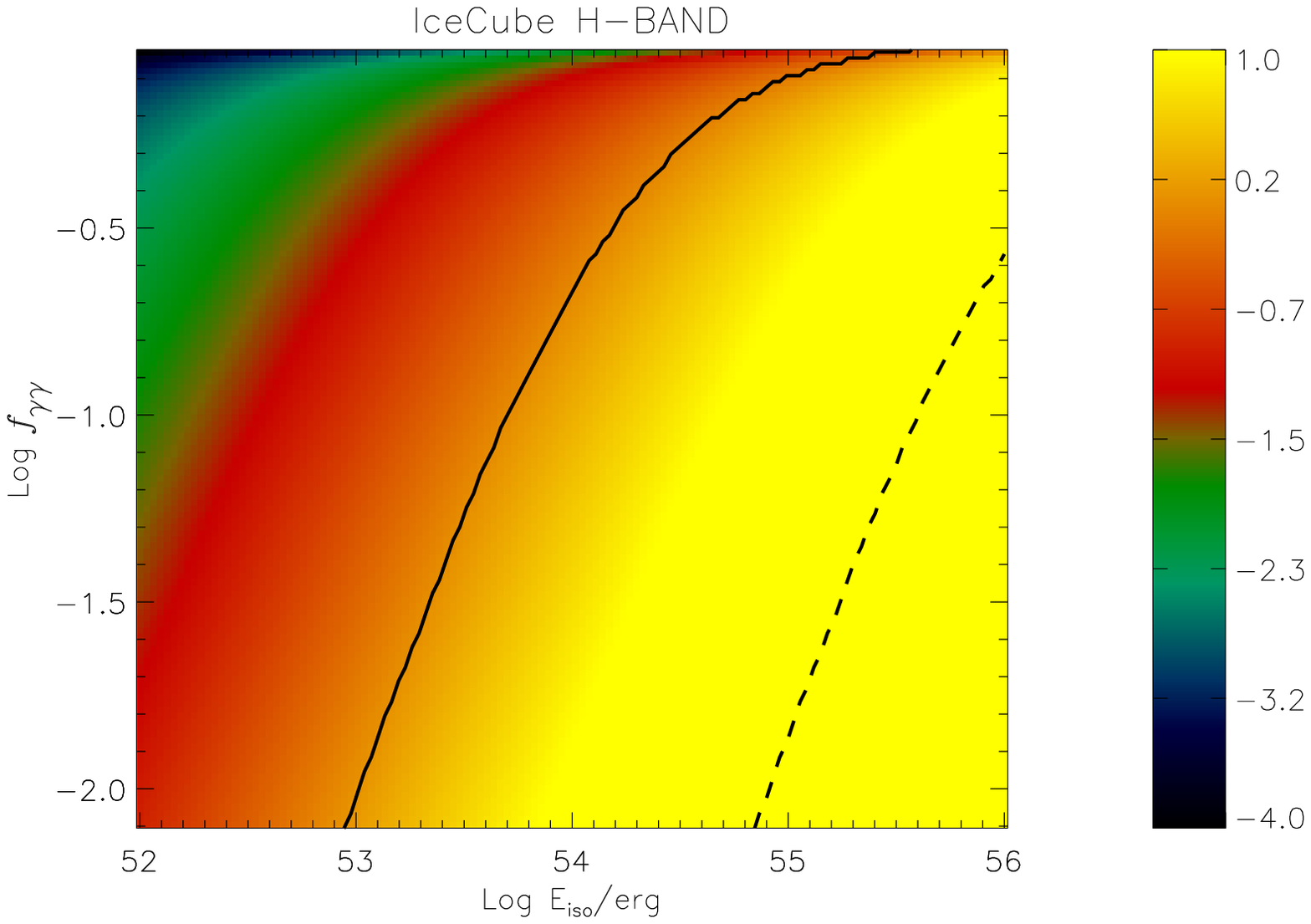,height=7.5cm}}
\caption{\footnotesize Same as Fig.~\ref{fig03} but for IceCube.}
\label{fig04}
\end{figure}

\end{document}